\documentclass[aps,prb,superscriptaddress,a4paper,twocolumn]{revtex4-1}
\usepackage{graphicx}
\usepackage{amsmath,amsfonts,amssymb,color}

\newcommand{\buenosaires}{Departamento de F\'isica and IFIBA, FCEN, Universidad de Buenos Aires, 
Ciudad Universitaria, Pab.\ I, C1428EHA Buenos Aires, Argentina}
\newcommand{\bayreuth}{Theoretische Physik III, Universit\"at Bayreuth, 95440 Bayreuth, Germany}

\begin{document}

\title{Ultrafast spin dynamics in II-VI diluted magnetic semiconductors with spin-orbit interaction}

\author{F.\ Ungar} 
\affiliation{\bayreuth}

\author{M.\ Cygorek}
\affiliation{\bayreuth}

\author{P.\ I.\ Tamborenea}
\affiliation{\buenosaires}
\affiliation{\bayreuth}

\author{V.\ M.\ Axt}
\affiliation{\bayreuth}

\date{\today}

\begin{abstract}
We study theoretically the ultrafast spin dynamics of II-VI diluted magnetic semiconductors
in the presence of spin-orbit interaction.
Our goal is to explore the interplay or competition between the exchange $sd$-coupling and
the spin-orbit interaction in both bulk and quantum well systems.
For bulk materials we concentrate on Zn$_{1-x}$Mn$_x$Se and take into account the Dresselhaus 
interaction, while for quantum wells we examine Hg$_{1-x-y}$Mn$_x$Cd$_y$Te systems 
with a strong Rashba coupling.
Our calculations were performed with a recently developed formalism which incorporates 
electronic correlations beyond mean-field theory originated from the exchange $sd$-coupling.
For both bulk and quasi-two-dimensional systems we find that, by varying the system parameters
within realistic ranges, both interactions can be chosen to play a dominant role
or to compete on an equal footing with each other.
The most notable effect of the spin-orbit interaction in both types of systems is the 
appearance of strong oscillations where the exchange $sd$-coupling by itself
only causes an exponential decay of the mean electronic spin components.
The mean-field approximation is also studied and it is interpreted analytically why it shows 
a strong suppression of the spin-orbit-induced dephasing of the spin component parallel 
to the Mn magnetic field.
\end{abstract}


\maketitle


\section{Introduction}

Diluted magnetic semiconductors (DMS) are multifunctional materials that combine the outstanding 
electronic and optical properties of semiconductors with highly controllable magnetic properties.
\cite{kos-gaj,xia-ge-cha}
With the prospect of spintronic applications of DMS in mind, much effort has focused 
recently on the study of ultrafast spin dynamics and control. 
\cite{kir-kim-ras,qi-xu-ste,has-kob-mun,cyw-sha,mor-her-man,kap-per-wic,thu-axt-win}
At the same time, spin-orbit interaction (SOI) effects have been intensely studied in non-magnetic
bulk and nanostructured semiconductors. 
\cite{win,zut-fab-das,fab-etal,dya,wu-jia-wen,tsy-zut}
The interplay between the exchange interaction characteristic of DMS and the more generic
SOI can lead to new possibilities for applications and basic research.
\cite{gna-nav,yan-cha-wu,val-mir,jia-zho-kor,mir-sho-ela,den-ave}
In particular, the spin-orbit torque effect in DMS has attracted much interest in recent years.
\cite{man-zha,gar-mac,hal-bra-tse,che-ove-liu,end-mat-ohn,fan-kur-wun,li-wan-doa,lin-wan-per}

In this article we explore theoretically this interplay by studying the ultrafast spin
dynamics of a non-equilibrium electron distribution in the conduction band of II-VI
Mn-doped semiconductors.
Our work is based on a microscopic density-matrix theory that models on a quantum-kinetic
level the spin precession and the spin transfer 
between electrons in the conduction band of such semiconductors and the manganese electrons,
and which accounts for exchange-induced correlations beyond the mean-field level and considers 
the localized character of the Mn spins.\cite{thu-axt} 
This recently developed formalism is quite general and can be computationally costly
to apply in some circumstances.
For this reason, in the present study we consider a particular situation which is 
nevertheless experimentally relevant and theoretically interesting:
the limit of high Mn density compared to the electron density, which is
normally realized in photoexcitation experiments.
In this particular regime we can apply a simplified formalism which captures the essential
physics that is relevant here and which reduces greatly the numerical effort.\cite{cyg-axt-15}

The purpose of our study is to determine under which conditions, if any, the spin-orbit
interaction mechanisms present in semiconductors can become relevant or even dominant
in the picosecond time-scale spin dynamics in DMS.
As will be seen here, for both bulk and quasi-two-dimensional systems, depending on
the choice of material parameters and excitation conditions, there can be a strong interplay 
or competition between the two types of interactions.
This rather unexplored combined effect between exchange and SOI in DMS
could lead in principle to new forms of spin control suitable for spintronic applications.

This article is organized as follows.
In Section \ref{sec:hamiltonian} we present the model Hamiltonian of the DMS with spin-orbit 
interaction and in Section \ref{sec:eq_motion} we review the equations of motion that describe 
the spin dynamics in the formalism adopted here.
In Sections \ref{sec:bulk} and \ref{sec:quantum_well} we present and discuss our results for bulk 
Zn$_{1-x}$Mn$_x$Se and for   Hg$_{1-x-y}$Mn$_x$Cd$_y$Te quantum wells, respectively.
Finally, we provide some concluding remarks.


\section{Quantum kinetic formalism}

\subsection{DMS Hamiltonian}
\label{sec:hamiltonian}

The theoretical model of DMS for our work includes the exchange $sd$-coupling 
between electrons in the conduction band and $d$-electrons of the doping Mn 
atoms and the SOI of conduction-band electrons expressed in the envelope-function approximation.
The Hamiltonian has the form
\begin{equation}
  H = H_0 + H_{sd} + H_{\text{SO}} \, ,
\label{eq:H_total}
\end{equation}
where $H_0 = \sum_i \mathbf{p}_i^2 / 2 m^{\ast}$, with conduction-band effective mass $m^{\ast}$,
and the Kondo-like Hamiltonian \cite{thu-axt} 
\begin{equation}
  H_{sd} = J_{sd} \sum_{iI} \mathbf{s}_i \cdot
                 \mathbf{S}_I\,\delta (\mathbf{r}_i - \mathbf{R}_I)
\label{H_sd_1}
\end{equation}
describes the coupling due to the exchange interaction between the 
conduction-band electrons and the Mn electrons.
The spin operator and position of the $I$-th Mn atom 
($i$-th conduction-band electron)
are denoted as $\mathbf{S}_I$ and $\mathbf R_I$ 
($\mathbf{s}_i$ and $\mathbf r_i$), respectively.
The coupling constant $J_{sd}$ is negative here, corresponding to a
ferromagnetic coupling.\cite{cib-sca}
In the present work the negative Land\'{e}-factor of Mn will always 
be combined with the negative sign of the coupling constant $J_{sd}$.
In addition, all spin variables will be considered dimensionless and
the coupling constant accordingly modified.

For bulk materials, the SOI Hamiltonian $H_{\text{SO}}$ 
of zincblende semiconductors is the Dresselhaus Hamiltonian \cite{dre}
\begin{equation}
H_{\text{D}} = \gamma_{\text{D}} \sum_i [\sigma_{i,x} k_{i,x} (k_{i,y}^2 - k_{i,z}^2) + 
                          \text{cyclic perm.}] \, ,  
\label{eq:dresselhaus}
\end{equation}
where $\boldsymbol{\sigma}$ is the vector of Pauli matrices and $\mathbf{k}$
is the operator $\mathbf{p}/\hbar$.
For quasi-two-dimensional systems, we consider asymmetric quantum wells which
display the Rashba SOI \cite{ras}
\begin{equation}
H_{R}= \alpha_{\text{R}} \sum_i (k_{i,y} \sigma_{i,x} - k_{i,x} \sigma_{i,y}) \, .
\label{eq:rashba}
\end{equation}
These effective spin-orbit couplings can be thought of as interactions of 
a spin with $\mathbf{k}$-dependent magnetic fields.


\subsection{Equations of motion}
\label{sec:eq_motion}

In Refs.\ [\onlinecite{thu-axt}], [\onlinecite{cyg-axt-15}], and [\onlinecite{cyg-axt}],
the Heisenberg equations of motion of the density matrix for DMS
without SOI were posed and analyzed in terms of a 
correlation hierarchy which includes averaging of the Mn-atom positions, 
thus rendering the problem spatially homogeneous.
In this work we follow that formalism and extend it in a simple fashion
in order to study the effects of the SOI on the electronic
spin degree of freedom.

When the number of Mn atoms ($N_{\text{Mn}}$) is much larger than the number of conduction-band 
electrons ($N_{\text{e}}$), i.e.\ in the limit $N_{\text{Mn}} \gg N_{\text{e}}$,
the quantum kinetic equations established in Ref.\ [\onlinecite{thu-axt}] can 
be significantly simplified.
This assumption can be easily fulfilled for intrinsic semiconductors
in which the Mn$^{2+}$ ions are incorporated isoelectronically, like in the case
of II-VI semiconductors.\cite{fur-kos}
Unlike the situation in, for example, III-V based DMS, where the Mn doping results in a large number of holes, 
in isoelectronically doped systems the density of free carriers is controlled solely by the
photoexcitation and thus can be kept much smaller than the Mn density
simply by using low laser intensities.
Here we consider electrons excited with typical narrow-band laser pulses 
with near-bandgap energies and low intensities.
Employing the approximation of low-electron density as compared to the Mn
doping density, we have developed a simplified 
formalism \cite{cyg-axt-15} based on the full model of Ref.\ [\onlinecite{thu-axt}] 
which allows a numerically efficient handling of electronic correlations.
Here we adopt the low electron-density limit and follow the formalism
of Ref.\ [\onlinecite{cyg-axt-15}].

In the regime $N_{\text{Mn}} \gg N_{\text{e}}$ the Mn density matrix can be considered
stationary and we take the $z$-axis along the mean Mn magnetization $\langle \mathbf{S} \rangle$.
The assumption of a stationary Mn density matrix has been numerically tested under conditions 
comparable with our present case in Refs.\ 
[\onlinecite{cyg-axt-15}],
[\onlinecite{cyg-axt}], 
[\onlinecite{thu-cyg-axt-kuh-b}], and 
[\onlinecite{thu-cyg-axt-kuh-a}].
We introduce a precession frequency for the conduction-band electrons in the 
effective magnetic field of the Mn atoms
\begin{equation}
  \omega_{\text{\tiny M}} = \frac{J_{sd}}{\hbar} n_{\text{Mn}} S \, ,
\label{omega_M}
\end{equation}
where $n_{\text{Mn}}$ is the Mn density and $S =|\langle \mathbf{S} \rangle|$,
with $0 \leq S \leq \frac{5}{2}$. 

\begin{figure*}[t!] 
\includegraphics{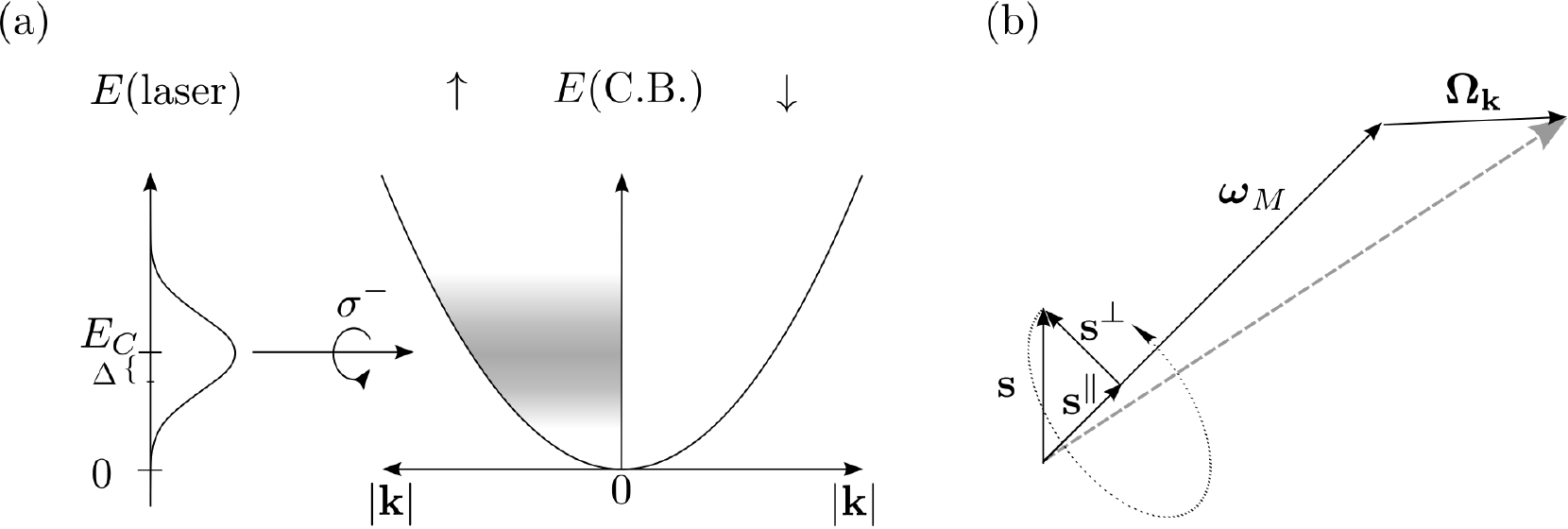}
\caption{
(a) Schematic representation of the conduction band (C.B.) and the spectrum of
the circularly-polarized Gaussian laser pulse that excites electrons from the valence
band to a Gaussian distribution of spin up electrons in the conduction band (centered at an
energy $E_C$ above the band edge and with standard deviation $\Delta$).
(b) The electron spin and its components $\langle \mathbf{s}_{\mathbf{k}}^{\perp}\rangle$ and 
$\langle s_{\mathbf{k}}^{\parallel} \rangle$, perpendicular and parallel to the Mn magnetic field
(or equivalently angular frequency $\boldsymbol{\omega}_\text{M}$), respectively. 
Also represented is  $\boldsymbol{\Omega}_{\mathbf{k}}$, the angular frequency associated with 
the $\mathbf{k}$-dependent spin-orbit effective magnetic field.
The electron spin precesses about $\boldsymbol{\omega}_\text{M}+\boldsymbol{\Omega}_{\mathbf{k}}$.
}
\label{fig:schematic}
\end{figure*}

We study the time evolution of the mean value of the spin operator associated with
the state with wave vector $\mathbf{k}$,
\begin{equation}
\langle \mathbf{s}_{\mathbf{k}} \rangle = 
    \sum_{\sigma \sigma'} \mathbf{s}_{\sigma \sigma'}
    \langle 
        c_{\sigma \mathbf{k}}^{\dagger}  
        c_{\sigma' \mathbf{k}}
    \rangle=
    (\langle \mathbf{s}_{\mathbf{k}}^{\perp} \rangle, \langle s_{\mathbf{k}}^{\parallel}\rangle),
\label{eq:spin_operator}
\end{equation}
where $\langle \mathbf{s}_{\mathbf{k}}^{\perp}\rangle$ and 
$\langle s_{\mathbf{k}}^{\parallel} \rangle$ 
are the mean spin components perpendicular and parallel to the mean Mn magnetization, 
respectively [see Fig.\ \ref{fig:schematic}(b)].
We will take as system variables $\langle \mathbf{s}_{\mathbf{k}}^{\perp}\rangle$
and the populations 
$n_{\mathbf{k}}^{\sigma} =    
    \langle 
        c_{\sigma \mathbf{k}}^{\dagger}  
        c_{\sigma \mathbf{k}}
    \rangle$.
The parallel mean spin can be obtained from the latter as
\begin{equation}
\langle s_{\mathbf{k}}^{\parallel} \rangle = \frac{1}{2} 
                                             \left(n_{\mathbf{k}}^{\uparrow}-
                                                   n_{\mathbf{k}}^{\downarrow}
                                             \right).
\end{equation}
Leaving aside for the moment the SOI, the time evolution of these variables induced by
$H_0$ and the $sd$-interaction is given by \cite{cyg-axt-15}
\begin{widetext}
\begin{equation}
\label{eq:n}
\left.
\frac{\partial}{\partial t} n_{\mathbf{k}}^{\uparrow / \downarrow} 
\right|_{\text{sd}}= 
  \sum_{\mathbf{k'}} 
  \left[
        \Re(G_{\omega_{\mathbf{k'}}}^{\omega_{\mathbf{k}}}) 
        \frac{b^{\parallel}}{2} 
        \left(  
           n_{\mathbf{k'}}^{\uparrow / \downarrow} - 
           n_{\mathbf{k}}^{\uparrow / \downarrow} 
        \right) 
        + \Re(G_{\omega_{\mathbf{k'}}}^{\omega_{\mathbf{k}} \pm \omega_\text{\tiny M}}) 
        \left( 
           b^{\pm} n_{\mathbf{k'}}^{\downarrow / \uparrow} - 
           b^{\mp} n_{\mathbf{k}}^{\uparrow / \downarrow} 
           \mp 2 b^{0} 
           n_{\mathbf{k}}^{\uparrow / \downarrow} 
           n_{\mathbf{k'}}^{\downarrow / \uparrow} 
        \right)
   \right] \, ,
\end{equation}
\begin{eqnarray}
\left.
\frac{\partial}{\partial t} \langle \mathbf{s}_{\mathbf{k}}^{\perp}\rangle 
\right|_{\text{sd}} &=&
-\sum_{\mathbf{k'}}
\left\{
   \left[ 
      \Re(G_{\omega_{\mathbf{k'}}}^{\omega_{\mathbf{k}} - \omega_\text{\tiny M}}) 
         \left(\frac{b^{+}}{2}\!-\!b^{0} n_{\mathbf{k'}}^{\uparrow}
         \right)
      +\Re(G_{\omega_{\mathbf{k'}}}^{\omega_{\mathbf{k}} + \omega_\text{\tiny M}}) 
         \left(\frac{b^{-}}{2}\!+\!b^{0} n_{\mathbf{k'}}^{\downarrow}
         \right)
   \right] 
   \langle\mathbf{s}_{\mathbf{k}}^{\perp}\rangle 
   + \Re(G_{\omega_{\mathbf{k'}}}^{\omega_{\mathbf{k}}}) 
   \frac{b^{\parallel}}{2}
   (\langle \mathbf{s}_{\mathbf{k'}}^{\perp} \rangle 
    +\langle \mathbf{s}_{\mathbf{k}}^{\perp} \rangle) \right\}   \nonumber \\
& + &
\left\{
   \omega_\text{\tiny M} 
 - \sum_{\mathbf{k'}} 
\left[ 
   \Im(G_{\omega_{\mathbf{k'}}}^{\omega_{\mathbf{k}} - \omega_\text{\tiny M}}) 
   \left( 
      \frac{b^{+}}{2} - b^{0} n_{\mathbf{k'}}^{\uparrow}
   \right) -  
   \Im(G_{\omega_{\mathbf{k'}}}^{\omega_{\mathbf{k}} + \omega_\text{\tiny M}}) 
   \left( 
      \frac{b^{-}}{2} + b^{0} n_{\mathbf{k'}}^{\downarrow}
   \right)
\right]
\right\}
\frac{\langle\mathbf{S}\rangle}{S} \times 
\langle \mathbf{s}_{\mathbf{k}}^{\perp} \rangle \, .
\label{eq:s}
\end{eqnarray}
\end{widetext}
The constants in Eqs.\ \eqref{eq:n} and \eqref{eq:s} depend only on the setting of the
Mn magnetization and are given by
$b^{\pm} = \langle {S^{\perp}}^2 \rangle \pm b^{0}$, 
$b^{0} = \langle S^{\parallel}\rangle/2$, 
$b^{\parallel} = \langle {S^{\parallel}}^2 \rangle$,
where
$ S^{\parallel} = \mathbf{S} \cdot \langle\mathbf{S}\rangle / S$, 
and
$\langle {S^{\perp}}^2 \rangle= \langle S^2 - {S^{\parallel}}^2 \rangle / 2$. 
We also called
$\omega_{\mathbf{k}} = E_{\mathbf{k}}/ \hbar = \hbar \mathbf{k}^2 / 2 m^{\ast}$.

The function $G_{\omega_{\mathbf{k'}}}^{\omega_{\mathbf{k}}}$ can be interpreted as a memory
function and has the form
\begin{eqnarray}
   G_{\omega_{\mathbf{k'}}}^{\omega_{\mathbf{k}}}(t) &=&  
      \frac{J_{sd}^2 \, n_{\text{\tiny Mn}}}{V\hbar^2} 
      \int_{-t}^{0} dt' e^{i(\omega_{\mathbf{k'}} - \omega_{\mathbf{k}})t'} \nonumber \\
      &\approx& \frac{J_{sd}^2 \, n_{\text{\tiny Mn}} }{V \hbar^2} 
      \pi \delta(\omega_{\mathbf{k'}} - \omega_{\mathbf{k}}),
\label{eq:G} 
\end{eqnarray}
where in the last step we neglected the imaginary part and the finite memory,
i.e.\ we applied a Markov limit which is a good approximation for not too large
values of $J_{sd}^2$ and excitations not too close to the band edge.\cite{thu-cyg-axt-kuh-a}

The spin-orbit Hamiltonians of Eqs.\ \eqref{eq:dresselhaus} or \eqref{eq:rashba} introduce,
to a first approximation, an additional $\mathbf{k}$-dependent spin precession.
If the contribution of a single electron with wave vector $\mathbf{k}$ to the spin-orbit 
Hamiltonian is written in the form
\begin{equation}
H_{\text{SO}} = \frac{\hbar}{2} \, \mathbf{\hat{\Omega}}_{\mathbf{k}} \cdot \boldsymbol{\sigma},
\end{equation}
then the mentioned spin precession is described by the Heisenberg equation of motion of 
the mean value of the spin operator introduced in Eq.\ (\ref{eq:spin_operator}),
\begin{equation}
\left.
\frac{\partial}{\partial t} \langle\mathbf{s}_{\mathbf{k}}\rangle 
\right|_{\text{SO}} =
    \mathbf{\Omega}_{\mathbf{k}} \times \langle\mathbf{s}_{\mathbf{k}}\rangle.
\end{equation}
Note that while $\mathbf{\hat{\Omega}}_{\mathbf{k}}$ is an operator, we introduced 
$\mathbf{\Omega}_{\mathbf{k}}$ as the corresponding regular vector
where $\mathbf{k}$ is interpreted simply as a wave vector and not as an operator as in 
Eqs.\ \eqref{eq:dresselhaus} and \eqref{eq:rashba}  [see Fig.\ \ref{fig:schematic}(b)].
In the present study we take into account the influence of the spin-orbit
interaction at this level, in order to elucidate how this added $\mathbf{k}$-dependent 
precession alters the quantum spin dynamics in bulk and quasi-two-dimensional DMS.


\section{Bulk Zn$_{1-x}$Mn$_x$Se}
\label{sec:bulk}

In this Section we present ultrafast spin dynamics results for bulk semiconductors.
For concreteness we focus on Zn$_{1-x}$Mn$_x$Se which is currently one of the best studied 
II-VI DMS, and as we will see, it can display an interesting interplay between exchange and SOI.
We first examine numerically and analytically the dephasing caused by the Dresselhaus 
spin-orbit coupling and then we proceed to calculate and analyze the full dynamics under 
the influence of both exchange coupling and SOI.

\subsection{Dresselhaus-induced dephasing}
\label{sec:dephasing}

\begin{figure}[h] 
\centerline{\includegraphics[width=3in]{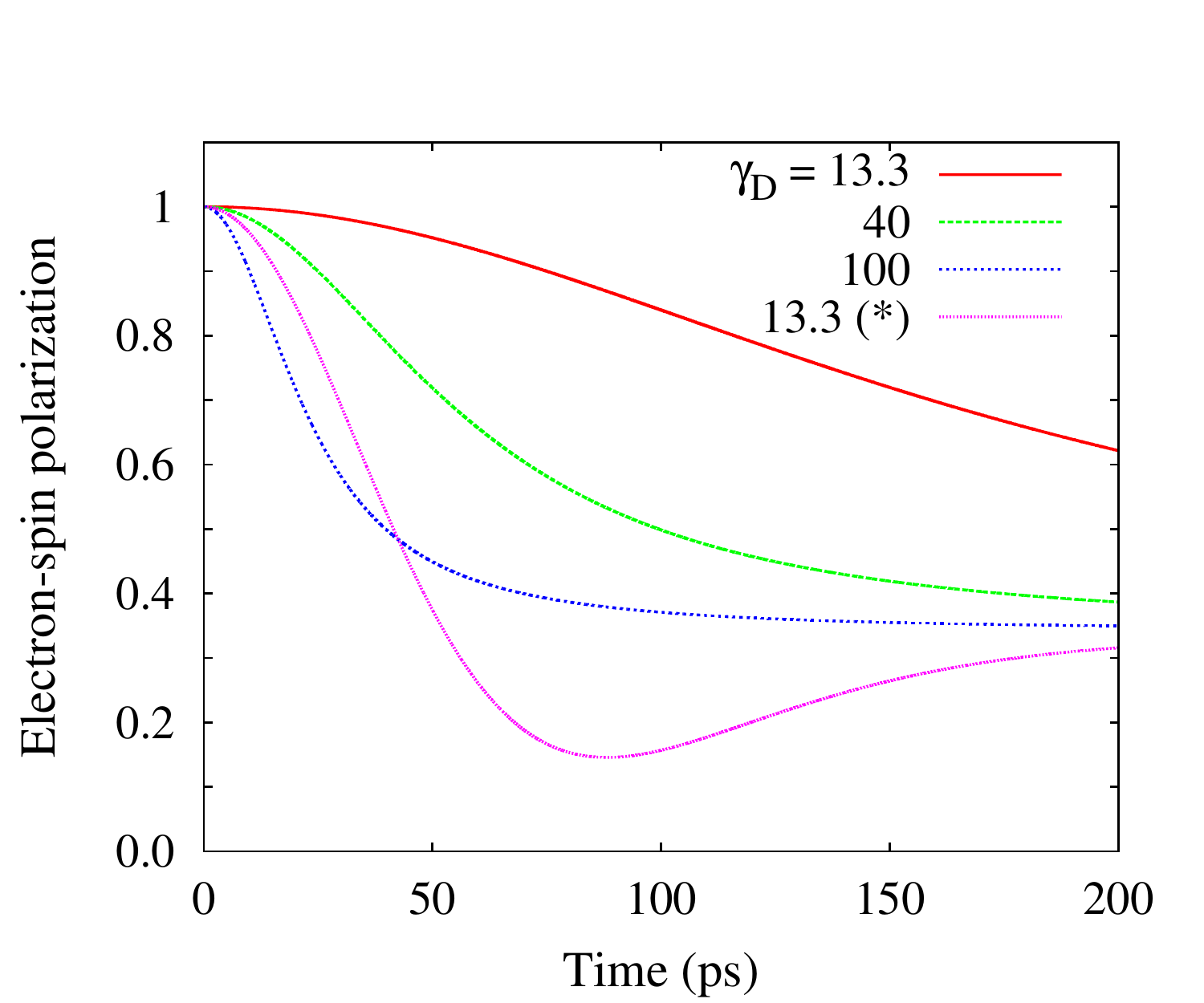}}
\caption{
Dephasing after isotropic Gaussian excitation of the spin-up band
(standard deviation $\Delta=$3 meV) without exchange $sd$-coupling in an 
effective Dresselhaus spin-orbit magnetic field with prefactor $\gamma_D$, specified in 
$\mbox{ps}^{-1} \mbox{nm}^{3}$.
Three of the curves correspond to a Gaussian excitation centered at the band edge
($E_C=0$), while the fourth, marked with (*), corresponds to a displacement of the excitation 
to $E_C=$10~meV above the band edge.
}
\label{fig:3D_SO_noMn}
\end{figure}

As mentioned in Sec.\ \ref{sec:hamiltonian}, the spin-orbit interaction in the 
envelope-function approximation plays the role of an effective 
$\mathbf{k}$-dependent magnetic field around which the electron spin precesses.
This spin precession in the case of an electron gas leads to global spin dephasing and decay, 
which is at the root of the D'yakonov-Perel spin-relaxation mechanism.\cite{dya-per}
As initial condition for the conduction-band electrons we assume a Gaussian distribution
caused by a pulsed optical excitation, similar to the one illustrated in 
Fig.\ \ref{fig:schematic}(a).
For the moment we consider a Gaussian distribution centered at $E_C=0$ (the band edge)
and later we will consider an excitation centered at $E_C=$10~meV, always with standard
deviation $\Delta =$3~meV.
We assume that the optical excitation populates only the spin-up conduction-band states
thanks to its appropriate circular polarization.
In Fig.\ \ref{fig:3D_SO_noMn} we plot the spin polarization,
$\langle s_z\rangle(t) = 2 N_{\text{e}}^{-1} \sum_{\mathbf{k}} \langle s_{\mathbf{k},z} \rangle(t)$
(normalized to 1), of the initially spin-up electron
population (the $z$-axis coincides with the main axis of the zincblende lattice)
in the conduction band versus time for different values of the Dresselhaus 
spin-orbit coupling constant $\gamma_D$.

The accepted standard value of 
$\gamma_{\text D}/\hbar = 13.3 \, \text{ps}^{-1} \text{nm}^3$ is included,\cite{win} 
and two artificially high values (40 and 100 $\text{ps}^{-1} \text{nm}^3$) are added 
to explore the tendencies of the decay behavior.
We use for the conduction-band effective mass of ZnSe the value
$m^{\ast}=0.134\,m_0$,\cite{yu-car} where $m_0$ is the bare electron mass.
The expected dephasing and decay mentioned above is clearly observed, with faster 
decay obtained for increasing SOI coupling constant.
Note that the decay, however, is not exponential from the beginning, but rather quadratic 
at short times.
Another interesting feature is that for an excitation 10 meV above the band edge the evolution
displays a non-monotonic behavior.
Below we shall indicate the origin of this incipiently oscillatory behavior.

The long-time limit of the spin polarization seen in Fig.\ \ref{fig:3D_SO_noMn}, 
which corresponds to the equilibrium distribution caused by the SOI effective field 
dephasing, is given by the value 1/3:
\begin{equation}
  \lim_{t \rightarrow\infty} \langle s_z\rangle(t) =: \langle s_{eq}\rangle 
                                        = \frac{1}{3} \langle s_z\rangle(t=0).
\label{eq:lim_t_infty_s}
\end{equation}
This equilibrium value can be understood analytically as follows.
The equation of motion for the spin under the SOI effective magnetic field, 
$\left. \frac{\partial}{\partial t} \langle\mathbf{s}_{\mathbf{k}}\rangle \right|_{\text{SO}} =
{\mathbf{\Omega}_{\mathbf{k}}} \times \langle\mathbf{s}_{\mathbf{k}}\rangle$,
can be cast in the matrix form
$\left. \frac{\partial}{\partial t} \langle\mathbf{s}_{\mathbf{k}}\rangle \right|_{\text{SO}} =
\mathbb{M}_{\mathbf{k}} \langle\mathbf{s}_{\mathbf{k}}\rangle$,
where
\begin{equation}
\mathbb{M}_{\mathbf{k}} = \left( \begin{array}{ccc}
                 0                     & -\Omega_{\mathbf{k},z} & \Omega_{\mathbf{k},y} \\
                 \Omega_{\mathbf{k},z} & 0                      & -\Omega_{\mathbf{k},x} \\
                -\Omega_{\mathbf{k},y} & \Omega_{\mathbf{k},x}  & 0\\
\end{array} \right),
\end{equation} 
and has the formal solution 
\begin{equation}
  \langle \mathbf{s}_{\mathbf{k}}\rangle(t) = 
  \exp( \mathbb{M}_{\mathbf{k}} t) \, \langle\mathbf{s}_{\mathbf{k}}\rangle(0).
\label{eq:evolution_op}
\end{equation}
The Taylor expansion of the matrix exponential can be simplified using that 
$\mathbb{M}_{\mathbf{k}}^3 = - \Omega_{\mathbf{k}}^2 \, \mathbb{M}_{\mathbf{k}}$ and
$\mathbb{M}_{\mathbf{k}}^4 = - \Omega_{\mathbf{k}}^2 \, \mathbb{M}_{\mathbf{k}}^2$,
with $\Omega_{\mathbf{k}}=|\boldsymbol{\Omega}_{\mathbf{k}}|$.
One obtains
\begin{equation}
  \exp(\mathbb{M}_{\mathbf{k}} t) = 1 + \sin(\Omega_{\mathbf{k}} t) \, 
                              \frac{\mathbb{M}_{\mathbf{k}}}{\Omega_{\mathbf{k}}} +
                              \left[1-\cos(\Omega_{\mathbf{k}} t)\right] 
                              \left(\frac{\mathbb{M}_{\mathbf{k}}}{\Omega_{\mathbf{k}}}\right)^2 .
\end{equation}
The diagonal elements of this matrix are given by
\begin{equation}
\exp(\mathbb{M}_{\mathbf{k}} t) |_{ii} = \frac{\Omega_{\mathbf{k},i}^2}{\Omega_{\mathbf{k}}^2} + 
                          \left( 1- \frac{\Omega_{\mathbf{k},i}^2}{\Omega_{\mathbf{k}}^2} \right) 
                          \cos(\Omega_{\mathbf{k}} t).
\label{eq:soi_oscillation}
\end{equation}
Assuming that initially only the $i$-th spin component is non-zero, from  
Eqs.\ (\ref{eq:evolution_op}) and (\ref{eq:soi_oscillation}) we obtain
$  \langle s_{\mathbf{k}i}\rangle(t) = 
  \exp( \mathbb{M}_{\mathbf{k}} t) |_{ii} \, \langle s_{\mathbf{k}i}\rangle(0).$
Thus, for large times $t$, this spin component, averaged over the isotropically occupied 
$\mathbf{k}$-states,
tends to $\langle s_i \rangle = \overline{\Omega_i^2 / \Omega^2} =1/3$ 
since the effective field is isotropic
(the bar denotes average over $\mathbf{k}$-states).

Note again that in Fig.\ \ref{fig:3D_SO_noMn} the curve corresponding to the excitation above
the band edge displays a non-monotonic behavior which is the precursor of an oscillation
that can be seen under stronger SOI.
These oscillations will be observed later in the quantum-well situation, and originate
from the cos-term in Eq.\ (\ref{eq:soi_oscillation}), appropriately averaged
over the occupied $\mathbf{k}$-states.


\subsection{Interplay between exchange and Dresselhaus interactions}
\label{sec:exchange-Dresselhaus}

Having verified the dephasing caused by the $\mathbf{k}$-dependent Dresselhaus effective magnetic
field, we now wish to study the interplay between the exchange $sd$- (sd) and Dresselhaus (D)
couplings.
The material parameters of Zn$_{1-x}$Mn$_x$Se related to the Mn doping used in our simulations
are as follows.
The exchange coupling constant of Zn$_{1-x}$Mn$_x$Se is 
$N_0 \alpha=260 \, \text{meV}$,\cite{twa-von-dem}
where $N_0$ is the number of unit cells per unit volume, 
and $\alpha=J_{sd}$ in our notation.
The lattice constant of ZnSe is 0.569~nm, the volume of the primitive
unit cell is 0.0455~nm$^3$, thus $N_0=22~\text{nm}^{-3}$, and then 
$J_{sd}\approx 12 \, \text{meV}\,\text{nm}^3$.
We assume a relatively low percentage of Mn doping of $0.3\%$
which gives a Mn density of  $6.6 \times 10^{-2}\,\text{nm}^{-3}$.
The density of photoexcited electrons is assumed to be $5 \times 10^{-5}\,\text{nm}^{-3}$,
i.e.\ three orders of magnitude lower than the Mn density.

\begin{figure}[ht]
\centerline{\includegraphics[width=3in]{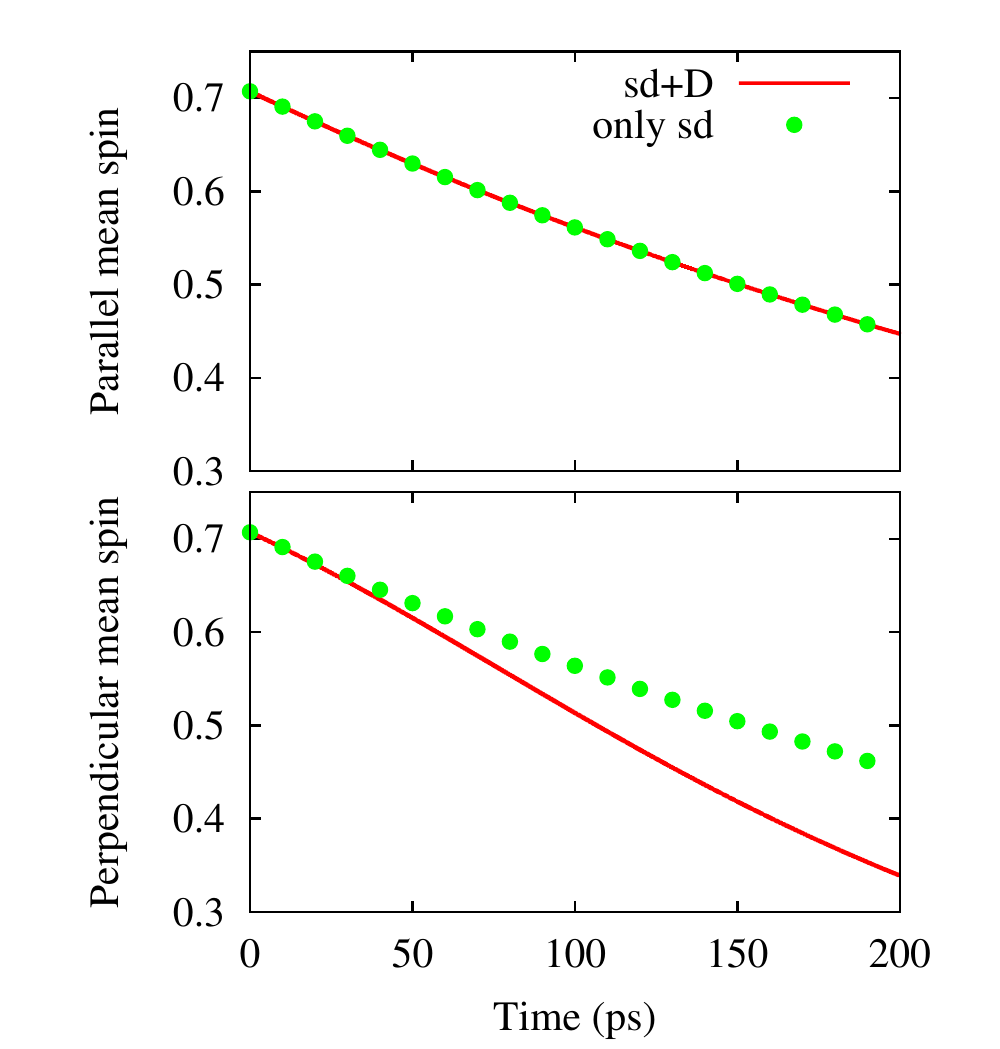}}
\caption{
Influence of the Dresselhaus spin-orbit coupling (D) on the spin dynamics in bulk 
Zn$_{1-x}$Mn$_x$Se with exchange $sd$-coupling (sd) for an initially
Gaussian electron occupation centered at the band edge with standard deviation $\Delta=$3 meV 
and initial spin-polarization rotated 45 degrees with respect to the z-axis.
The Mn concentration is $x_{\text{Mn}} = 0.3\%$ and the net Mn magnetization $S = 0.5$.
Red solid lines correspond to the full calculation (sd+D) and green dotted lines to the
calculation leaving out the Dresselhaus coupling (only sd).
}
\label{fig:bulk_bandedge}
\end{figure}

We first consider a Gaussian distribution for the conduction-band electrons centered 
at the band edge, and take an average Mn magnetization of $S = 0.5$.
The Mn magnetization can be simply tuned by applying an external magnetic field in the
desired direction and waiting for the Mn spin to reach its thermal equilibrium.
Thus, we envision an experiment where the magnetic field is turned off before the pump 
laser pulse arrives. 
Note that the Mn spin-lattice relaxation time is of the order of $0.1 \, \mu\text{s}$ 
\cite{koe-mer-yak} which suffices to carry out the ensuing optical excitation
experiment studied here under almost constant Mn magnetization. 
Figure \ref{fig:bulk_bandedge} shows the time evolution of the parallel, $\langle s_z\rangle(t)$,
and perpendicular,
$|\langle \mathbf{s}^{\perp} \rangle(t)| = 
  2 N_{\text{e}}^{-1} |\sum_{\mathbf{k}} \langle \mathbf{s}_{\mathbf{k}}^{\perp} \rangle(t)|$,
mean spin components. 
From now on we use only the realistic value
$\gamma_{\text D}/\hbar = 13.3 \, \text{ps}^{-1} \text{nm}^3$
for the Dresselhaus constant and for concreteness we take the initial spin-polarization 
rotated 45 degrees with respect to the z-axis.
The specific choice for this angle is not very relevant, but it is important to set it
to a value different from zero in order to have spin precession about the Mn field.
Since the Dresselhaus Hamiltonian is cubic in the wave vector, we expect it to have
a relatively weak effect, as compared to the exchange coupling, on electrons populating 
low-energy states around the band edge, and Fig.\ \ref{fig:bulk_bandedge} confirms this expectation.
Indeed, we see that for the parallel spin component the presence of the Dresselhaus coupling 
does not modify the dynamics noticeably [the red-solid line (sd+D) and the green dots (only sd) 
are superimposed].
For the perpendicular components there is a noticeable difference, but the two curves 
are still qualitatively similar.
We have checked that if the Mn concentration and/or the Mn magnetization are increased 
the effect of the spin-orbit coupling becomes rapidly negligible also for the perpendicular
spin component.
Roughly speaking, the exchange $sd$-coupling can be thought of as causing two main effects: 
a spin precession about the mean Mn magnetization and spin transfer between conduction-band
and Mn electrons.
On the other hand, as seen above, the Dresselhaus spin-orbit Hamiltonian, 
by providing a $\mathbf{k}$-dependent effective magnetic field, induces a global dephasing 
in the electron population.
The decay seen in both spin components in Fig.\ \ref{fig:bulk_bandedge} is thus a result of
both exchange-induced spin transfer and spin-orbit dephasing, but the former dominates
the dynamics for the chosen set of parameters.

\begin{figure}[ht]
\centerline{\includegraphics[width=3.in]{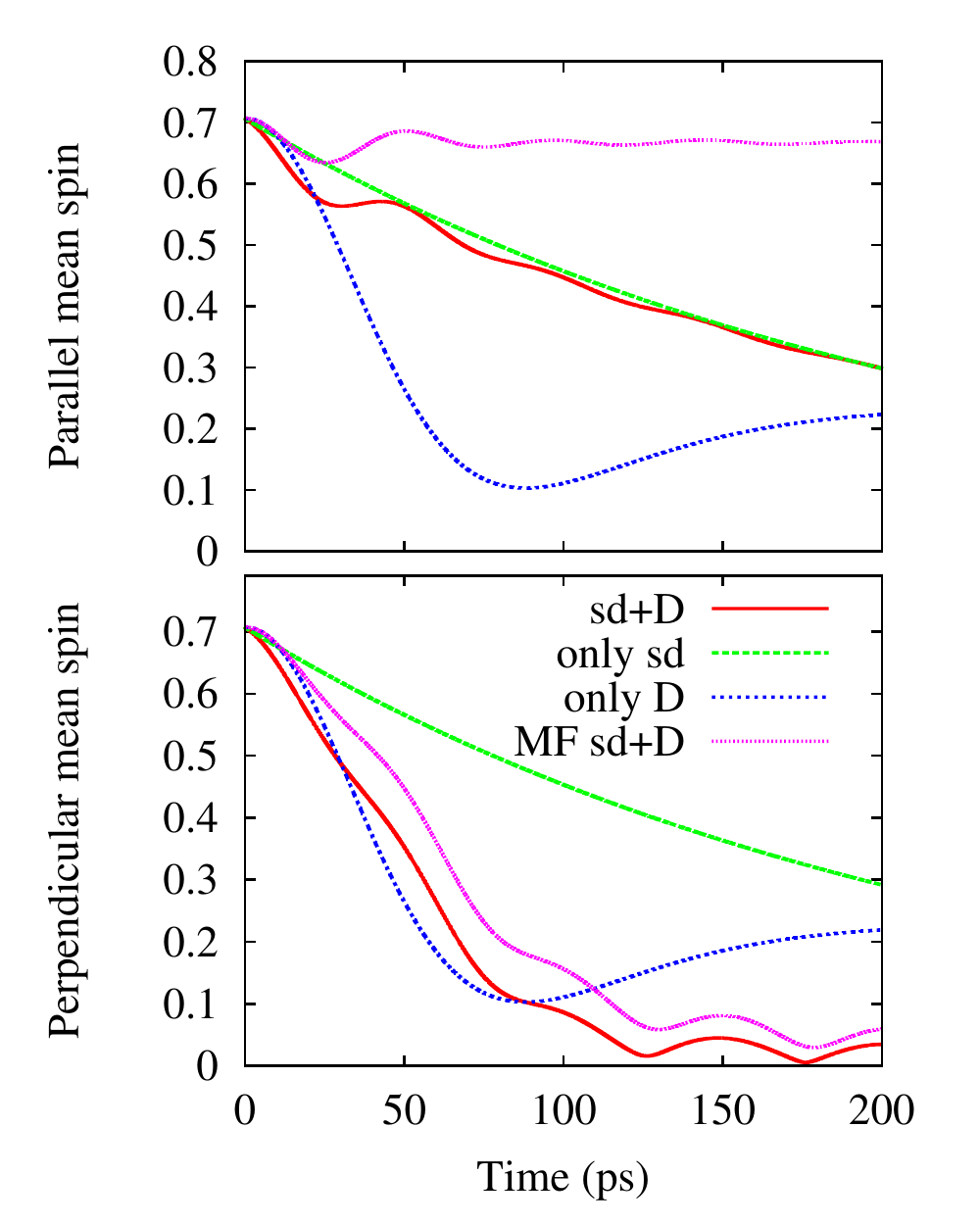}}
\caption{
Influence of the Dresselhaus spin-orbit coupling (D) on the spin dynamics in bulk 
Zn$_{1-x}$Mn$_x$Se with exchange $sd$-coupling (sd) for an initially 
Gaussian electron occupation centered at $E_C=$10 meV above the band edge with 
standard deviation $\Delta=$3 meV and initial spin-polarization rotated 45 degrees 
with respect to the z-axis.
The Mn concentration is $x_{\text{Mn}} = 0.3\%$ and the net Mn magnetization $S = 0.1$.
Red solid lines: full calculation with sd+D; 
green long-dashed lines: only sd;
blue short-dashed: only D; 
pink dotted lines: mean-field approximation with sd+D.
}
\label{fig:bulk_10meV}
\end{figure}

This raises the question of whether a parameter regime can be reached experimentally in which 
the dephasing caused by the spin-orbit effective field has a considerable influence on or even 
dominates the spin dynamics.
As mentioned above, shifting the optical excitation away from the band edge to higher 
k-values should enhance the effect of the SOI on the spin dynamics.
Furthermore, the influence of the exchange $sd$-coupling can be reduced by lowering both
the Mn concentration and/or the average Mn magnetization.
Thus, in Fig.\ \ref{fig:bulk_10meV} we show the time evolution of the parallel and perpendicular 
spin components like in Fig.\ \ref{fig:bulk_bandedge}, but centering the Gaussian occupation
10 meV above the band edge and reducing the Mn magnetization to $S = 0.1$.
The Mn doping is kept at $x_{\text{Mn}} = 0.3\%$ as before, and for the conduction-band 
electrons we choose again an initial spin orientation rotated 45 degrees away from the z-axis.
In the parallel spin component there is now a noticeable difference between the full
calculation (sd+D) (red solid line) and the sd-only case (green long-dashed line).
A qualitatively new feature is that the combination of $sd$- and Dresselhaus couplings 
now produces not only a decay but also oscillations, revealing a combined spin precession.
In the perpendicular component the spin-orbit coupling has now an enormous effect,
greatly accelerating the decay and causing superimposed oscillations.
The oscillations seen in Fig.\ \ref{fig:bulk_10meV} have a frequency close to the precession
frequency associated with the mean Mn magnetic field
($\boldsymbol{\omega}_{\text{\tiny M}} = \omega_{\text{\tiny M}} \mathbf{\hat{z}}$), 
$\omega_{\text{\tiny M}} = 0.124 \, \text{Thz}$  (period $T_{\text M}=50.7 \, \text{ps}$).
We come back to this issue after discussing the mean-field approximation which we
now introduce.

It is interesting to elucidate whether a similar spin dynamics would also be obtained
in a simpler scenario combining the Dresselhaus SOI with a constant magnetic field
of appropriate strength. 
(This type of problem has been studied recently from the point of view of impurity 
entanglement \cite{met-dil-egg} and spin relaxation \cite{zho-yu-wu}.)
We can readily answer this question by intentionally leaving the correlation terms 
out of the equations of motion 
[keeping in the RHS of Eq.\ \eqref{eq:s} only the first term of the second line] 
thus reverting to a mean-field approximation, which for a given Mn magnetization
is equivalent to adding a constant magnetic field.
The result is given by the pink dotted lines in Fig.\ \ref{fig:bulk_10meV}.
For the perpendicular component we see that the mean-field calculation resembles the full
one, although there is a clearly distinguishable difference between them.
On the other hand, both results are far away from the $sd$-only result, and we have to
conclude that in this sense the mean-field approximation does capture an important part 
of the interplay between the exchange and spin-orbit couplings.
For the parallel component the mean-field approximation radically modifies the dynamics.
We see here that when the $sd$ correlations are removed the spin-orbit dephasing is not
capable by itself of inducing a decay in the presence of the spin precession about
the Mn magnetization.
In other words, the longitudinal component does not decay since its Dresselhaus dephasing is
in a sense prevented by the 'naked' (without exchange-induced correlations) precession about 
the Mn magnetization.
To confirm this point we show in Fig.\ \ref{fig:bulk_10meV} the spin dynamics with only
the Dresselhaus SOI (no $sd$-coupling) with blue dotted lines.
These curves show the strong decay induced by spin-orbit dephasing in the absence of both
the spin precession and the spin transfer caused by the exchange coupling.

The origin and frequency of the oscillations mentioned above, which appear when both
interactions are present, and in both the full and mean-field calculations, can be 
interpreted with the help of Eq.\ (\ref{eq:soi_oscillation}).
For a given $\mathbf{k}$-state the precession frequency is now 
$\mathbf{\Omega} \equiv \mathbf{\Omega}_{\mathbf{k}} + \omega_{\text{\tiny M}} \mathbf{z}$.
In the limit $\omega_{\text{\tiny M}} \gg \Omega_{\mathbf{k}}$ we can
assume that $\Omega_z \approx \Omega \approx \omega_{\text{\tiny M}}$, 
and using Eqs.\ (\ref{eq:evolution_op}) and (\ref{eq:soi_oscillation}) we obtain 
$\langle s_z \rangle(t) = \langle s_z \rangle(0)$.
This argument applies to every $\mathbf{k}$-state and thus can be extended to the
whole electron population.
Then, the precession about the spin-orbit effective magnetic field of the longitudinal component 
is suppressed by the dominant precession about the Mn magnetic field, a feature that
can be seen clearly in the mean-field result of Fig.\ \ref{fig:bulk_10meV}.
If the spin-orbit angular frequency is not completely neglected we obtain oscillations
in the parallel component with frequency 
$|\mathbf{\Omega}_{\mathbf{k}} + \boldsymbol{\omega}_{\text{\tiny M}}|$ 
and small amplitude proportional
to $1-(\Omega_{\mathbf{k},z}+\omega_{\text{\tiny M}})^2/ 
|\mathbf{\Omega}_{\mathbf{k}} + \boldsymbol{\omega}_{\text{\tiny M}}|^2$,
as seen in  Fig.\ \ref{fig:bulk_10meV}.
We have verified that increasing the Dresselhaus coupling increases 
the amplitude of the oscillations (not shown here).
Oscillations of the same frequency are also present in the perpendicular spin component.


\section{Hg$_{1-x-y}$Mn$_x$Cd$_y$Te quantum wells}
\label{sec:quantum_well}

We now turn to the study of the influence of the spin-orbit coupling in II-VI semiconductor 
quantum wells.
In this case the SOI that we consider is the Rashba coupling (R), which is
present when the quantum-well confinement lacks inversion symmetry.
As explained in the Introduction, the role of the spin-orbit coupling is conceptually similar 
in bulk and in quantum wells, since in both cases it can be thought of as a $\mathbf{k}$-dependent
Zeeman Hamiltonian which induces global dephasing in an electron gas.
However, quantum wells offer greater flexibility to control the SOI and also
display high electron mobilities in high quality modulation-doped samples.
High mobilities amount to longer momentum-scattering times and therefore to more coherent
quantum dynamics.

In line with the bulk studies discussed above, we first tested the spin dynamics
in Zn$_{1-x}$Mn$_x$Se quantum wells.
For realistic parameters for this material, it turned out that the Rashba coupling was too weak
to modify the dynamics driven by the exchange $sd$-coupling.
The root of this difficulty seems to be the large bandgap (about 2.8 eV) of ZnSe, which results 
in a small Rashba coupling constant.
Thus for the quantum well calculations we looked for a family of materials with stronger and 
more controllable Rashba interaction.
Hg$_{1-x}$Mn$_{x}$Te is a good candidate since the energy gap $E_g$ of this ternary 
compound depends strongly on the Mn concentration,\cite{fur} going to zero at $x \leq 6.5\%$, 
while its spin-orbit valence-band splitting $\Delta$ is insensitive to it.\cite{kos}
This interesting combination leads to flexible spin-orbit properties, 
which are generally controlled by the ratio $\Delta/E_g$.
By choosing a Mn concentration slightly above $6.5\%$ we can select a very low energy gap, 
which leads in turn to a strong Rashba coupling.\cite{dea-lar-bas}
However, this lower limit for the  Mn concentration is still too high and leads again 
to a completely dominant exchange interaction even for as low a Mn magnetization of $S =0.01$ 
(with $sd$-coupling induced spin relaxation times below 5 ps).
This drawback can be overcome by considering instead the compound Hg$_{1-x-y}$Mn$_{x}$Cd$_{y}$Te 
in which the non-magnetic Cd atoms replace some of the Mn dopants.
This change maintains the gap tunability via the doping fraction $x+y$ giving full flexibility 
regarding the concentration of magnetic ions.\cite{kos}
The Rashba coupling constant can be calculated with the expression \cite{dea-lar-bas,gna-nav}
\begin{equation}
 \alpha_R = \frac{\hbar^2}{2 m^{\ast}} \frac{\Delta}{E_g}
            \frac{2 E_g+\Delta}{(E_g+\Delta)(3E_g+2\Delta)}
            \frac{V_{\text{qw}}}{d} \, .
\end{equation}

\medskip\noindent
We work with the effective mass of HgTe, $m^{\ast}=0.093\,m_0$,\cite{yu-car}
and take the spin-orbit valence-band splitting as $\Delta=1.08\,\text{eV}$.\cite{zhu-xia}
Assuming $E_g=300 \,\text{meV}$, 
a quantum-well width $d=200 \, \text{\AA}$,
and a potential energy drop of $V_{\text{qw}} =50 \,\text{meV}$ across the quantum well, 
we obtain $\alpha_R = 4.87 \, \text{meV} \, \text{nm}$
($\alpha_R  / \hbar = 7.4 \, \text{ps}^{-1} \text{nm}$).
Note that for ZnSe one obtains $\alpha_R = 0.015 \, \text{meV} \, \text{nm}$
($\alpha_R / \hbar = 0.023 \, \text{ps}^{-1} \text{nm}$),
a very low value which leads to negligible spin-orbit effects, as mentioned before.
For the exchange $sd$-coupling constant of HgMnTe we take 
$N_0 \alpha = 400 \, \text{meV}$,\cite{liu-buh-nov} and
the lattice constant of $0.645\,\text{nm}$ leads to
$J_{sd} = 26.8 \, \text{meV}\,\text{nm}^3$.
We keep the previous Mn concentration of $x=0.3\%$.

\begin{figure}[ht]
\centerline{\includegraphics[width=3.in]{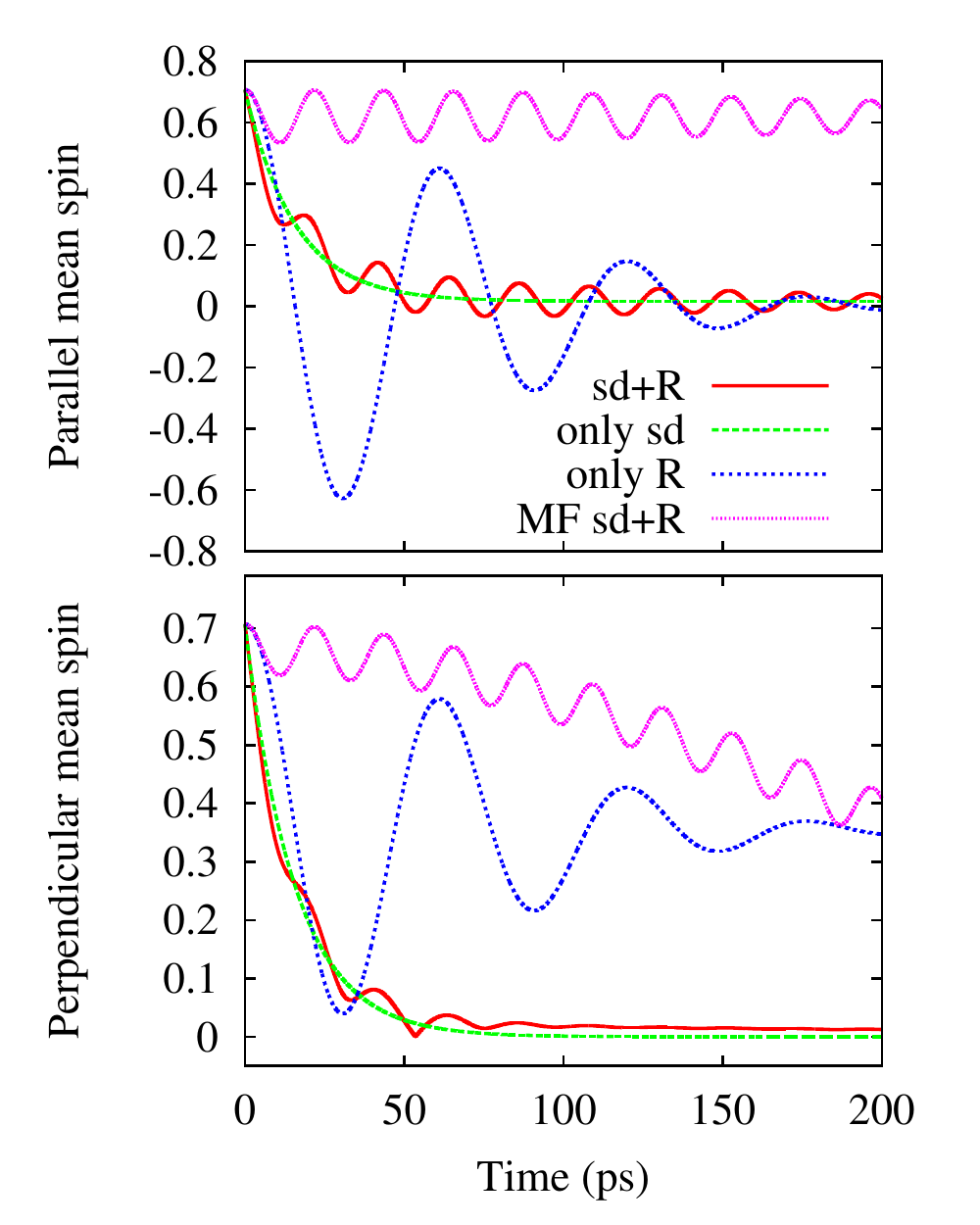}}
\caption{
Influence of the Rashba spin-orbit coupling (R) on the spin dynamics in a 
Hg$_{1-x-y}$Mn$_x$Cd$_y$Te quantum well with exchange $sd$-coupling (sd) 
for an initially Gaussian electron occupation centered 10 meV above the band edge 
with standard deviation of 3 meV and initial spin-polarization rotated 45 degrees 
with respect to the z-axis.
The Mn concentration is $x_{\text{Mn}} = 0.3\%$ and the net Mn magnetization $S = 0.1$.
Red solid lines: full calculation with sd+R; 
green long-dashed lines: only sd;
blue short-dashed lines: only R; 
pink dotted lines: mean-field approximation with sd+R.
}
\label{fig:QW}
\end{figure}

In Fig.\ \ref{fig:QW} we show the time evolution of the parallel and perpendicular spin
components for quantum wells, where now the parallel component corresponds to the growth 
direction of the quantum well ($z$-axis).
The Gaussian occupation is centered 10 meV above the band edge and we consider 
a Mn magnetization of $S = 0.1$.
The initial spin orientation is rotated 45 degrees away from the z-axis.

Figure \ref{fig:QW} shows that while the sd-only curve (green long-dashed line) follows 
the usual exponential decay, the full dynamics with sd+R displays clear oscillations 
in both components.
We have verified that the amplitude of these oscillations increases with increasing
Rashba coefficient, which in turn is obtained by lowering the energy gap.
The pink-dotted-lines in Fig.\ \ref{fig:QW} show the mean-field approximation results 
with both sd+R.
Here the decay seen in the perpendicular component is due to the Rashba-induced dephasing 
since exchange correlations are absent.
The parallel component maintains an approximately constant mean value in agreement with
the analysis done in the previous Section, and shows a slight decrease of the oscillation 
amplitude due to the dephasing induced by the Rashba SOI.
We verified that this amplitude reduction is accelerated by increasing the Rashba coupling 
constant.
The blue-short-dashed lines show the evolution of the spin with only the Rashba interaction
present (no sd).
Here we see the full-fledged oscillations that had been anticipated in the discussion of
Fig.\ \ref{fig:3D_SO_noMn}.
These oscillations are the collective result of the individual spin precessions about 
the effective $\mathbf{k}$-dependent Rashba magnetic field.
To the best of our knowledge, an analytical expression or a simple interpretation for 
the frequency of these oscillations is not currently available.
This frequency depends on many factors such as the Rashba coefficient, the electron density, 
and the electronic distribution (which in our simulations is determined by the mean value and 
the standard deviation of the initial Gaussian population).
We have checked numerically that there is a roughly linear dependence of this frequency
on the Rashba coefficient for an excitation 10 meV above the band edge.

It is unexpected and noteworthy that, in the quantum-well case, 
the addition of the spin-orbit interaction to the DMS produces strong oscillations 
while at the same time leaves fairly unchanged the decay rate for our parameters, 
as can be seen in Fig.\  \ref{fig:QW} (``only sd'' versus ``sd+R'' curves).
Finally, we point out that the main qualitative difference between the results shown in 
Fig.\ \ref{fig:bulk_10meV} for bulk and Fig.\ \ref{fig:QW} for a quantum well is that, 
for the perpendicular spin component in the quantum well, the sd-only curve stays near 
the full result and the MF curve moves strongly away, while the opposite behavior occurs 
in bulk.


\section{Conclusion}

We studied theoretically the combined effects of the exchange $sd$-coupling and the spin-orbit
interaction in II-VI diluted magnetic semiconductors (DMS), both in bulk and in quantum wells.
Although our results can be considered generally valid in zincblende semiconductor systems, 
we focused on particular materials that show clearly the interplay between the two mechanisms: 
Zn$_{1-x}$Mn$_x$Se for bulk and Hg$_{1-x-y}$Mn$_x$Cd$_y$Te for quantum wells.
In our calculations we employed a recently developed formalism which incorporates
electronic correlations originated from the exchange $sd$-coupling.
The main conclusion of our study is that for both bulk and quasi-two-dimensional systems there
can be a strong interplay or competition between the two types of interactions, leading to
experimentally detectable signatures (for example in time-resolved Faraday and Kerr 
rotation experiments) of the spin-orbit interaction in DMS.
In bulk we find that the spin components parallel and perpendicular to the net Mn magnetization
have rather different responses to the presence of the spin-orbit (Dresselhaus) interaction,
the latter being much more affected by it.
Indeed, coherent oscillations---with a frequency reflecting the precession around a 
combination of the Mn magnetization and the Dresselhaus field---develop as a consequence 
of the interplay between the two interactions, which are completely absent when the 
exchange interaction dominates.
In addition, the decay rate is greatly enhanced for the perpendicular component
by the presence of the Dresselhaus interaction in the studied regime.
Regarding quantum wells, we find that the exchange interaction tends to be more dominant over
the spin-orbit interaction (Rashba coupling in this case), which led us to consider a family 
of materials with large valence-band-splitting spin-orbit constant and tunable energy gap.
For these DMS materials we obtained again a strong effect of the spin-orbit interaction, 
manifesting itself in the occurence of oscillations which are not seen when the exchange
interaction acts alone.
Remarkably, even though the combination of exchange and spin-orbit interaction leads to
clearly visible oscillations, the decay of the spin polarization is practically unaffected
by the presence of the Rashba interaction.
These signatures should be detectable experimentally in pump-and-probe experiments.
Finally, for both bulk and quantum wells we find that in the mean-field approximation treatment 
of the exchange interaction there is a strong suppression of the spin-orbit-induced dephasing 
of the spin component parallel to the Mn magnetic field.
The studied interplay between the spin-orbit interaction and the exchange coupling could 
improve spin control and thereby facilitate potential spintronic applications of DMS.

\acknowledgments
We gratefully acknowledge the financial support of the Deutsche Forschungsgemeinschaft
through grant No.\ AX17/9-1.
Financial support was also received from the Universidad de Buenos Aires, project UBACyT 
2011-2014 No. 20020100100741, and from CONICET, project PIP 11220110100091.


\end{document}